\documentclass[iop]{emulateapj}
\usepackage{amsmath}
\usepackage{natbib}

\submitted{Accepted by the Astrophysical Journal}

\usepackage{soul}
\usepackage{color}
\usepackage{textcomp}

\usepackage{graphicx}
\usepackage{hyperref}

\newcommand{\fu}{4U~1630$-$47}
\newcommand{\swift}{\textit{Swift}}
\newcommand{\maxi}{\textit{MAXI}}
\newcommand{\chandra}{\textit{Chandra}}
\newcommand{\rxte}{\textit{RXTE}}
\newcommand{\asec}{\ensuremath{''}}
\newcommand{\wsim}{\ensuremath{\sim}}

\begin{document}


\title{Dust scattering halo of 4U~1630$-$47 observed with Chandra and Swift: New constraints on the source distance}

\author{E. Kalemci\altaffilmark{1},
	T. J. Maccarone\altaffilmark{2},
        J. A. Tomsick\altaffilmark{3}
}

\altaffiltext{1}{Faculty of Engineering and Natural Sciences, Sabanc\i\ University, Orhanl\i-Tuzla, 34956, Istanbul, Turkey}

\altaffiltext{2}{Department of Physics \& Astronomy, Texas Tech University, Box 41051, Lubbock TX, 79409-1051 USA}

\altaffiltext{3}{Space Sciences Laboratory, 7 Gauss Way, University of
California, Berkeley, CA, 94720-7450, USA}


\begin{abstract}

We have observed the Galactic black hole transient \fu\ during the decay of its 2016 outburst with \chandra\ and \swift\ to investigate the properties of the dust scattering halo created by the source. The scattering halo shows a structure that includes a bright ring between 80\arcsec\ and 240\arcsec\ surrounding the source, and a continuous distribution beyond 250\arcsec. An analysis of the $^{12}$CO $J=1-0$ map and spectrum in the line of sight to the source indicate that a molecular cloud with a radial velocity of -79 km s$^{-1}$ (denoted MC $-79$) is the main scattering body that creates the bright ring. We found additional clouds in the line of sight, calculated their kinematic distances and resolved the well known "near" and "far" distance ambiguity for most of the clouds. At the favored far distance estimate of MC $-79$, the modeling of the surface brightness profile results in a distance to \fu\ of 11.5 $\pm$ 0.3 kpc. If MC $-79$ is at the near distance, then \fu\ is at 4.7 $\pm$ 0.3 kpc. Future \chandra, \swift, and sub-mm radio observations not only can resolve this ambiguity, but also would provide information regarding properties of dust and distribution of all molecular clouds along the line of sight. Using the results of this study we also discuss the nature of this source and the reasons for the anomalously low soft state observation observed during the 2010 decay.  

\end{abstract}



\section{Introduction}\label{sec:intro}

Galactic black hole transients (GBHTs) are sources that allow studies of strong gravity, accretion and outflows, as well as the surroundings by using them as luminous probes. GBHTs undergo outbursts where the luminosity can change by a factor of $\sim 10^{7}$. The multiwavelength observational properties change during outbursts, generally obeying a general trend in their evolution \citep{Belloni10jp}. Outbursts start in the hard state with a power-law that dominates the $\sim$1-20 keV spectrum, and stays hard as the mass accretion rate from the secondary increases. Compact, conical jets and strong X-ray variability (up to 50\% rms) are observed in the hard state. A transition to the soft state occurs near the peak of the outburst \citep{Miyamoto95, Maccarone03_a} during which blackbody like emission from a thin accretion disk dominates the X-ray spectrum. Finally, as the outburst decays, the source makes a transition back to the low hard state. The multi-wavelength studies of GBHTs during the outburst decays show a common trend of evolution as well \citep{Kalemci13, Dincer14}. One such trend that is still puzzling is the fact that all sources make a transition to the hard state at a relatively narrow range of 1\%-4\% Eddington luminosity ($L_{Edd}$) as discussed first in \cite{Maccarone03_b}. More importantly this narrow state transition luminosity range is not unique to GBHTs, and valid for neutron stars as well \citep{Maccarone03_b}.

\subsection{4U~1630$-$47}

\begin{figure}
\plotone{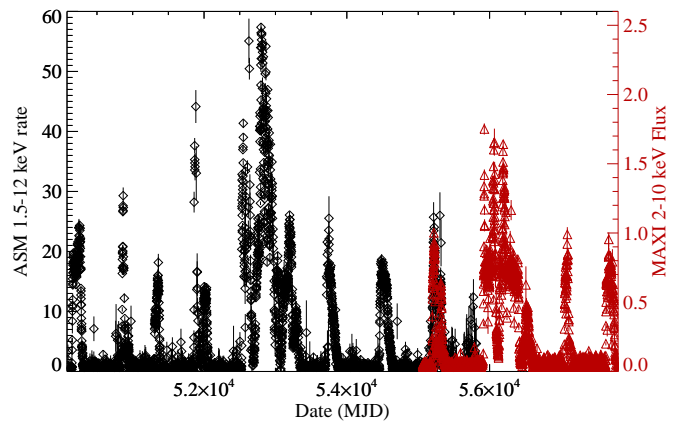}
\caption{\label{fig:maxiasm}
\rxte\ ASM (black, 1.5-12 keV band) and \maxi\ GSC (red, 4-10 keV band) light curves of \fu. 
}
\end{figure}

Among many GBHTs studied, one source stands out as peculiar: \fu. This source has had a large number of outbursts allowing us to compare the source spectral evolution in different outbursts (see Figure~\ref{fig:maxiasm}.) 

\epsscale{1.25}
\epsscale{1.0}

At least one well-studied outburst follows the typical hysteresis pattern described in \citep{Tomsick00}. However, most other outbursts have shown very different behavior \citep{Tomsick05, Abe05}. In the 2010 outburst decay, using \textit{Neil Gehrels Swift Observatory} (hereafter \swift) XRT, a very low luminosity soft state has been discovered, at  $L/L_{Edd}$=0.03 $M^{-1}_{10}$\% (with bolometric correction, distance taken as 10 kpc) where $M^{-1}_{10}$ is the mass of the black hole in units of 10 $M_{\odot}$ \citep[][hereafter T14]{Tomsick14}. Apart from the erratic outburst behavior, another peculiarity is the apparent baryonic content of the jet. The baryonic jet interpretation is based on modelling the iron line features observed simultaneously with radio originated from a local jet  \citep{DiazTrigo13}. We must note that models without presence of baryonic jets can explain the data as well \citep{Wang16}, and an even a simpler explanation could be that the radio emission is unrelated to iron line features, because it is caused by the interaction of a previously launched jet with a thick ISM \citep{Neilsen14}.

Why does \fu\ behave differently than other sources? First of all, this source is behind a high, and varying hydrogen column density \citep[][T14]{Augusteijn01}. Second, the source is surrounded by a dust scattering halo (DSH) as evidenced by archival \chandra\ and \swift\ observations (see \S\ref{sec:results}). Our hypothesis is that some of the peculiarities of \fu\ can be explained with the presence of a local dust cloud, and to test this hypothesis we analyzed archival \chandra\ data together with recently obtained data taken at the end of the 2016 outburst utilizing our \chandra\ observing program and \swift\ TOO observations.  While it is well known that dust scattering affects X-ray spectra of X-ray binaries \citep[][and references therein]{Smith16}, to the best of our knowledge, the effect of DSH on spectral properties at very low luminosity levels have not been studied extensively.

\subsection{Dust scattering halos}
\label{subsec:DSHintro}

Studies with dust scattering halos \citep{Overbeck65, Mathis91, Predehl95}, especially in high mass X-ray binaries, have been used to understand spectral variations during eclipses \citep{Audley06}, the physical properties of the dust grains and their distribution of grains along the line of sight \citep{Corrales15, Xiang11}, X-ray extinction \citep{Predehl95, Corrales16}, and to calculate the distance to the source by relating changes in DSH profile to flares or long term X-ray evolution of the sources  \citep{Trumper73, Thompson09, Xiang11}. 

\begin{figure}
\plotone{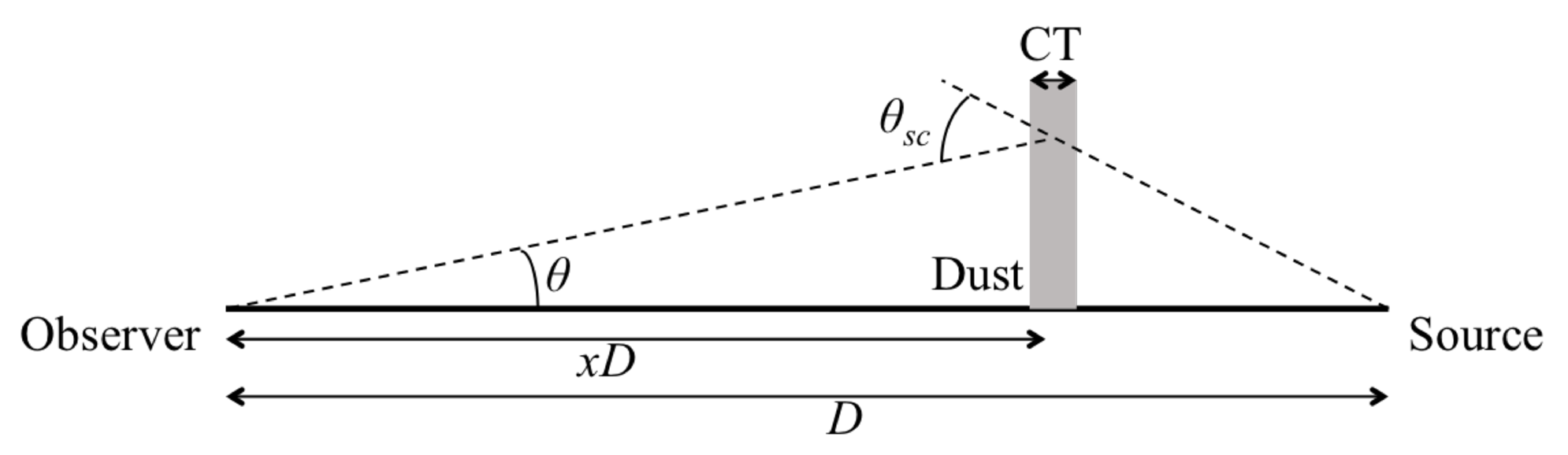}
\caption{\label{fig:dshgeom}
Cartoon depicting formation of dust scattering clouds. The source distance is given by $D$, and the distance to the dust cloud from the observer is given by $xD$. $\theta$ is the observed scattering angle, $\theta_{sc}$ is the physical scattering angle and $CT$ is the cloud thickness.
}
\vspace{0.5 pt}
\end{figure}

If a source exhibits outbursts followed by a long period of quiescence, the dust scattered emission takes the form of discrete rings. The main reason for this is that the most of the dust along the line of sight to a source would be in molecular clouds, and each ring is due to a single cloud producing delayed scattered emission as X-rays traverse a longer path than the X-rays directly observed by the telescope \citep[][and references therein]{Heinz16}. A cartoon image of the dust scattering geometry for single scattered emission is given in Figure~\ref{fig:dshgeom}.

\epsscale{1.2}
\epsscale{1.0}

As the behaviour of \fu, and the nature of \swift\ and \chandra\ observations are similar to those of V404 Cyg in 2015 outburst, we followed a similar methodology to analyze our data as \cite{Heinz16}. The relevant formulation is derived from simple geometrical arguments;

The time delay of arrival of scattered photons ($\Delta t$) from a source at distance $D$ and from a dust cloud at $xD$ is given by

\begin{equation}
	\label{eqn:deltat}
	\Delta t = \frac{xD\theta^{2}}{2c(1-x)}
\end{equation}

where $c$ is the speed of light. The observed intensity of a ring due to single scattering from a cloud (with thickness $CT$ much less than $D$ and outburst timescale shorter or comparable to $\Delta t$) is 

\begin{equation}
	\label{eqn:flux}
	I_{\nu,r} = N_{H,r} \frac{d\sigma_{sc,E}}{d\Omega} \frac{F_{\nu} (t=t_{obs}-\Delta t)}{(1-x)^{2}} \exp{[-\sigma_{ph,E}\sum\limits_{i=1}^r N_{H, r}]}
\end{equation}	

where  $\frac{d\sigma_{sc}}{d\Omega}$ is the differential dust scattering cross section per hydrogen atom, $F_{\nu}(t)$ is the flux of outburst at time $t$, $\sigma_{ph,E}$ is the total photo-electric cross section at energy $E$, $N_{H,r}$ is the hydrogen column density corresponding to the dust cloud and the sum is taken over all clouds (including a uniform dust distribution continuum) in between the cloud in question and the observer. The scattering cross section is a strong function of both the scattering angle and energy and depends on the dust distribution along the line of sight. In this work we used a simplified functional form of the cross section as:
  
\begin{equation}
	\label{eqn:crosssec}
	 \frac{d\sigma_{sc,E}}{d\Omega} \sim C \bigg(\frac{\theta_{sc}}{1000\arcsec}\bigg)^{-\alpha} \bigg(\frac{E}{1keV}\bigg)^{-\beta}
\end{equation}

where $C$ is a normalization (scattering cross section per hydrogen atom at 1000\arcsec\ and 1 keV), $\alpha \sim 3 - 4$ and $\beta \sim 3-4$ \citep{Draine03} and the physical scattering angle is simply

\begin{equation}
	\label{eqn:scatang}
	\theta_{sc} = \frac{\theta}{1-x}
\end{equation}

as can be deduced from Figure~\ref{fig:dshgeom}.

\begin{table*}
\centering
\caption{Observations used in the analysis} 
\begin{tabular}{ccccc}
\hline \hline
\multicolumn{5}{c}{Chandra Observations} \\
Date (MJD) & obsid & exposure & state  & Notes \\
55728.5 & 12530 & 19.3 ks & -  &Norma Arm for background, no DSH, no point source \\
57789.4 & 19004 & 39.2 & Hard &  Faint mode, ring DSH \\ \hline
\multicolumn{5}{c}{Swift XRT Observations} \\
55392.5 &  00031224006 & 4.6  & an. soft   & PC mode, DSH, T14 \\
57769.8 &  00031224046 & 2.3  &  Hard & WT mode \\
57771.8 &  00031224047 & 2.6  &  Hard &  PC mode, ring DSH \\
57773.2 &  00031224048 & 2.2  &  Hard & PC mode, ring DSH \\
57777.9 &  00031224050 & 2.0  &  Hard & PC mode, ring DSH \\
57781.9 &  00031224052 & 1.9  & Hard & PC mode, ring DSH\\
57783.8 &  00031224053 & 1.7  &  Hard & PC mode, ring DSH, no point source \\
57787.0 &  00031224055 & 2.3  &  Hard & PC,  mode, ring DSH, no point source \\
57789.0 &  00031224056 & 2.9  &  Hard & PC mode, ring DSH, no point source \\ \hline

\end{tabular}
\label{table:obs}
\end{table*}


\section{Observations and Analysis}\label{sec:obs}

We have utilized several pointed observations with \chandra\ and \swift\ as summarized in Table~\ref{table:obs}. Apart from pointed observations, we have also used \maxi\ GSC \citep{Matsuoka09} and \swift\ BAT to understand the spectral evolution of \fu\ during the 2016 outburst. The \maxi\  and BAT light curves, hardness ratios and the dates of \swift\ and \chandra\  observations can be seen in Figure~\ref{fig:maxibat}.The pointed observations took place when the source flux was already too low to be detected over the background by the all sky monitors. 

\subsection{Swift observations and point-source analysis}

During the 2016 outburst decay of \fu, we have obtained several TOO observations with \swift. The first observation on MJD~57769.8 was in the Windowed Timing (WT) and the rest of observations (from MJD~57771.8 to 57789.0) are in Photon Counting (PC) mode. After detection of the DSH in the \swift\ observations we have extended the analysis to all \swift\ PC mode observations in the archive.

For all observations for which \fu\ was detected, we determined the point source flux and spectral parameters using High Energy Astrophysics Software (HEASOFT) v6.20. We first produced photon event lists and exposure maps using \emph{xrtpipeline}. We extracted the photons from a circle with a 20-pixel (47\arcsec) radius centered on \fu\ for the source spectrum, and used a source-free and DSH-free region for the background spectrum. We created the exposure-map corrected ancillary response matrix is created using  \emph{xrtmkarf}, and selected the appropriate response matrix from the calibration database (\textit{CALDB}).

\subsection{Chandra observations and point-source analysis}

We have investigated all \chandra\ archival observations to search for extended emission around \fu. We eliminated Continuous Clocking mode observations as they did not produce an image of the DSH. While the DSH is apparent in four \chandra\ observations (obsids 13714 through 13717), due to high count rates the observations were taken in a sub-chip and graded as well, which made it difficult to analyze the extended emission. Similarly, obsid 15511 uses a sub-chip, and is extremely piled up, making it difficult to use. Obsid 15524 taken on MJD~56439.8 is mildly piled up and shows the extent of the DSH at 2\% Eddington luminosity. The analysis of the DSH profile of this observation will be done in a separate work. Here, we only utilized the archival observation of the Norma Arm (obsid 12530) taken on MJD~55728.5 to assess the possibility of extended background emission from other sources when \fu\ was in quiescence. 

As the outburst decayed at the end of 2016 we have triggered our \chandra\ observation and conducted a single observation on MJD 57789.4. Since the \swift\ flux evolution indicated very low count rates, we asked to remove the gratings and only obtained the image in the ACIS-S chip.

 We used \chandra\ Interactive Analysis of Observations ($CIAO$) v4.9 tools for the analysis of the Advanced CCD Imaging Spectrometer \citep[ACIS,][]{Garmire03}.  We created event and aspect solution files using \textit{chandra\_repro}. Using standard procedures we cleaned the event list from flares (which resulted in removing less than 1\% of events). For obsid 19004, the source spectrum was extracted from within a radius of 8\arcsec\, and the background was extracted within an annulus of 12\arcsec\-50\arcsec\ chosen to be inside the DSH ring. The spectrum is created using the $specextract$ tool of $CIAO$.

\epsscale{1.1}
\begin{figure*}
\plotone{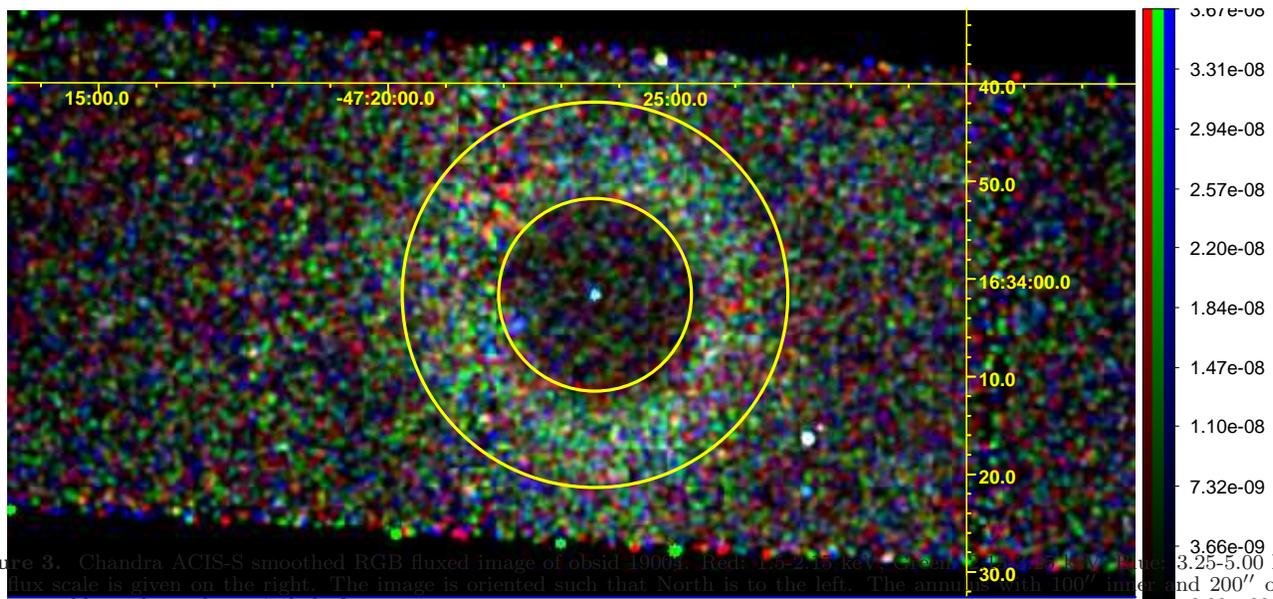}
\caption{\label{fig:chanring}
Chandra ACIS-S smoothed RGB fluxed image of obsid 19004. Red: 1.5-2.15 keV, Green: 2.15-3.25 keV, Blue: 3.25-5.00 keV.  The flux scale is given on the right. The image is oriented such that North is to the left. The annulus with 100\arcsec\ inner and 200\arcsec\ outer radius roughly encloses the main halo feature.
}
\vspace{0.5 pt}
\end{figure*}
\epsscale{1.0}

\vspace{1 pt}
 
\epsscale{1.25}
\begin{figure}
\plotone{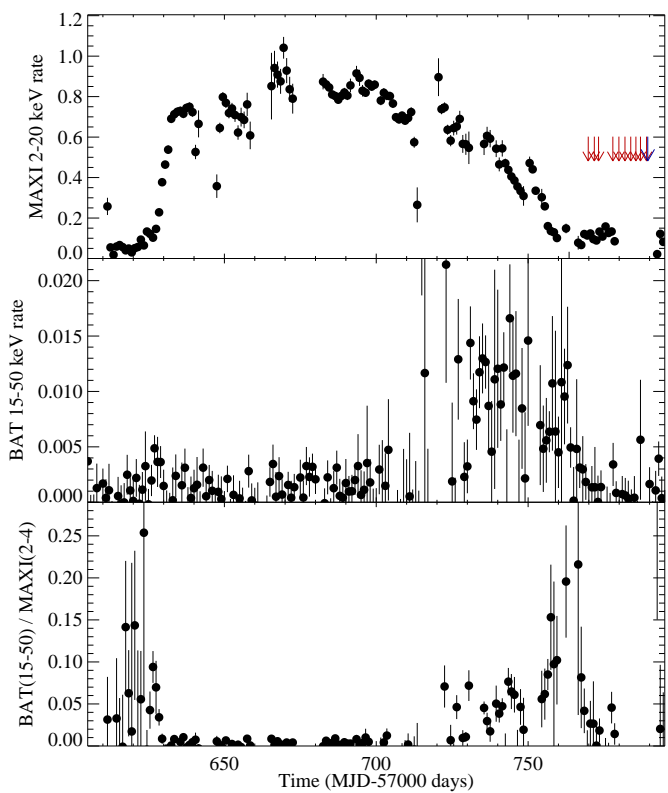}
\caption{\label{fig:maxibat}
Top: \maxi\ count rates in 2-20 keV range for 2016 outburst of \fu. Red arrows indicate times of pointed \swift\ observations, and the thick blue arrow shows the time of the \chandra\ observation. Middle: \swift\ BAT 15-50 keV count rate. Bottom: Ratio of the 15-50 keV BAT rate to 2-4 keV \maxi\ rate to indicate hardness.
}
\end{figure}
\epsscale{1.0}

\subsection{Analysis methods related to the DSH}
\label{subsec:dshmethod}

To calculate surface brightness profile (SBP) using the \chandra\ data we followed these steps:

\begin{itemize}
	\item determine the energy range in which the DSH is present over a background, find 3 energy bands with approximately equal counts in the SBP.
	\item compute fluxed images in these 3 energy bands (a composite image is given in Figure~\ref{fig:chanring}).
	\item find point sources in the image using $wavdetect$ \citep{Freeman02}, remove flux in circular areas around the source positions that would correspond to 99\% of full energy deposition in the PSF. 
	\item Divide the fluxed images into equal angular thickness circular regions, find the total flux and area of each region that is observable in ACIS-S. Apply a exposure-map cut-off of 15\% to calculate flux and areas.
	\item Apply a background correction.
	\item Divide corrected fluxes in each ring by the angular area of each ring to obtain the final SBP in each band. Area of the rings are corrected for detected point sources and chip boundaries.
\end{itemize}	

For the background correction, we first created fluxed images for each CCD using the standard blank-sky fields as described in the CIAO thread and \cite{Hickox06}. The normalization of the background fields are calculated by matching hard counts above 10 keV in each CCD separately with the corresponding blank-sky background files. Using these images we obtained background SBP profiles in each energy band for each CCD. The random errors in each bin are calculated by finding total counts in each ring and assuming Poisson statistics. Several factors affect the systematic errors in the process of obtaining surface brightness profiles such as image and spectral uniformity\footnote{\url{http://space.mit.edu/ASC/docs/expmap_intro.pdf}}. There are additional errors arising from \textit{blanksky} background subtraction as well. We added 3\% systematic errors to the entire SBP to take those effects into account. 

For the \swift\ profiles, the procedure was slightly different. We first made an energy cut in the event file, (1.5 - 5 keV) and divide the resultant image by the exposure map. We then divided the image into equal angular thickness circular regions from the source to the edge of the chip. For each ring we calculated the average energy of events, and the average off-axis angle of the ring and used the parametrization in \textit{CALDB} file $swxvign20010101v001.fits$ to obtain a vignetting correction in each ring. The vignetting correction amounts to $<$ 5\% inside 250\arcsec\ and becomes as large as 25\% at 600\arcsec\ \citep{Moretti09}. To obtain fluxed images in units of photons cm$^{-2}$ s$^{-1}$ we divided each count with the effective area corresponding to the energy of the photon. Finally we divided the flux in each ring with the angular area of the rings to obtain the SBP. The background is determined from the SBP at radii $> 600\asec$. Vignetting and background correction are the main sources of systematic error in \swift\ SBPs. We also added 3\% systematic errors to the \swift\ SBPs.

To calculate PSF profiles, we used \textit{Chart} simulations for \chandra, and the King function discussed in \cite{Moretti05} for \swift\ XRT.


\epsscale{1.1}
\begin{figure*}
\plottwo{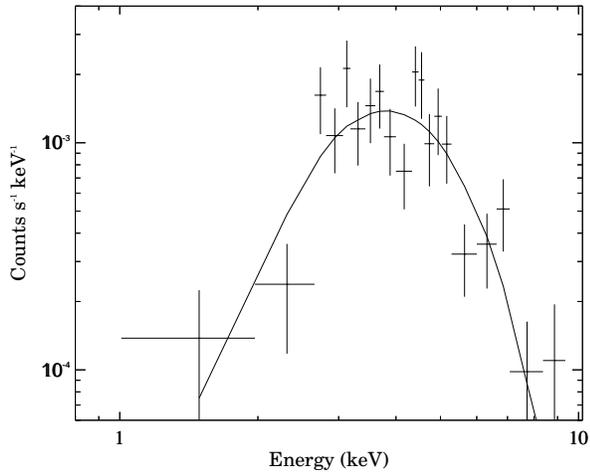}{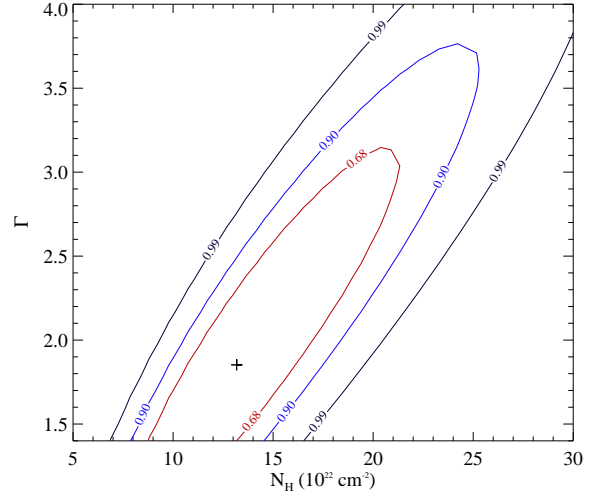}
\caption{\label{fig:pospec}
Left: \chandra\ point source spectrum and fit. Right: $\Delta$C-statistics contours of 90\%, 95\%, and 99\% confidence levels for the photon index and $N_{H}$. 
}
\end{figure*}
\epsscale{1}

\section{Results}\label{sec:results}

\subsection{Evolution of the outburst}

The evolution of soft and hard X-ray counts, and the hardness ratio is given in Figure~\ref{fig:maxibat}. The outburst starts on \wsim\ MJD~57620 and quickly rises, reaching a plateau of 0.8 cts/s in \maxi\ 2-20 keV rate in a timescale of 10 days. Like some of the previous outbursts, the initial hard state is not clearly observed. In fact, the BAT/\maxi\ ratio, and the \maxi\ evolution in different energy bands indicate that the hardness did not vary significantly throughout the outburst until around MJD~57730. After this date, an increase in the BAT count rate indicates a hardening, and the \swift\ observations after MJD~57750 are all consistent with being in the hard state.

\subsection{\chandra\ point source spectrum}

The \chandra\ point source spectrum within 8\arcsec\ source region and 12\arcsec\ -- 25\arcsec\ background region  is shown in Figure~\ref{fig:pospec}.  The spectrum is fitted with a model $tbabs \times\ power$-$law$ with abundances of \cite{Wilms00} and cross sections of \cite{Verner96}. Cash statistics are used for the fit \citep{Cash79}. This gives 2-10 keV unabsorbed flux of $2.6 \times 10^{-13}$ ergs$^{-1}$ cm$^{-2}$ s$^{-1}$, a photon index $\Gamma$ of $1.85 \pm 0.50$ with an $N_{H}$ of $1.3 \times 10^{23}$ cm$^{-2}$. Using a fiducial mass of 10 $M_{\odot}$ and a distance of \wsim\ 11.5 kpc as discussed in this work, the observed Eddington-scaled luminosity in 1-200 keV band is $\sim1.3 \times10^{-5}$. 

The photon index is critically dependent on the $N_{H}$ at these count rates. To determine how $\Gamma$ and $N_{H}$ are correlated we calculated $\Delta$C-statistics contours of 90\%, 95\%, and 99\% confidence levels using $steppar$ in $XSPEC$. The results, shown in Figure~\ref{fig:pospec}, indicate that for typical hard state spectrum the $N_{H}$ should be greater than $7 \times 10^{22}$ cm$^{-2}$, consistent with previous reports of the hydrogen column density (T14). For the photon indices commonly observed at these luminosity levels \citep{Plotkin13}, the $N_{H} > 10^{23}$ cm$^{-2}$.

\subsection{\chandra\ DSH profile results}

Figure~\ref{fig:chanring} shows the RGB composite fluxed image of the region around \fu\ obtained by the \chandra\ ACIS-S. The energy ranges that we used are 1.5 keV - 2.25 keV, 2.25 keV - 3.15 keV and 3.15 keV to 5 keV. The energy ranges are chosen to have approximately equal number of counts in each band. A DSH ring and a point source can be clearly observed.

\begin{figure}
\plotone{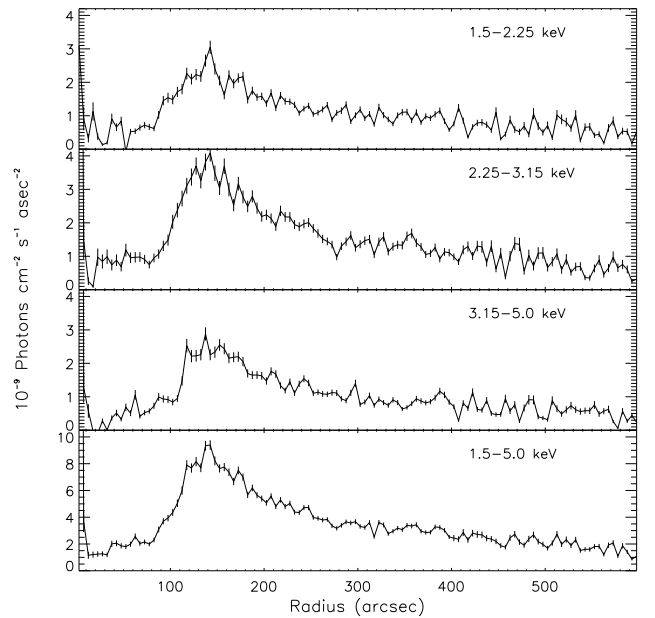}
\caption{\label{fig:SBPdifen}
Surface brightness profile of the \chandra\ obsid 19004 in 3 energy bands (1.5-2.25 keV, 2.25-3.15 keV, 3.15-5 keV) and the total 1.5-5 keV energy band. 
}
\end{figure}

After going through the procedure discussed in \S\ref{subsec:dshmethod}, we obtained SBP for the three energy bands used in the composite image, as well as the SBP for the full 1.5 keV to 5 keV range as shown in Figure~\ref{fig:SBPdifen}. 

\epsscale{1.25}
\epsscale{1.0}

\subsection{\swift\ DSH profile results}

\begin{figure*}
\plottwo{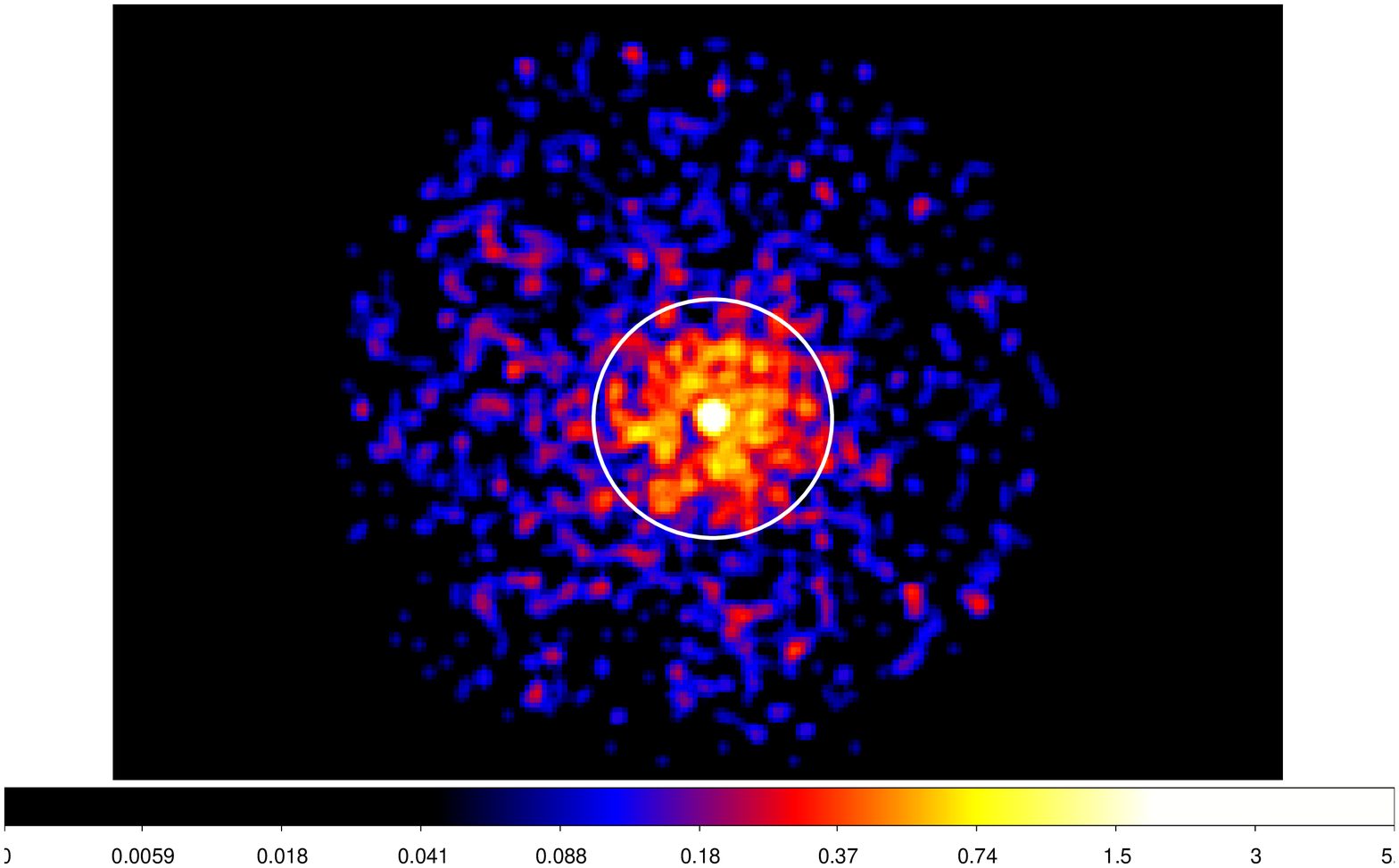}{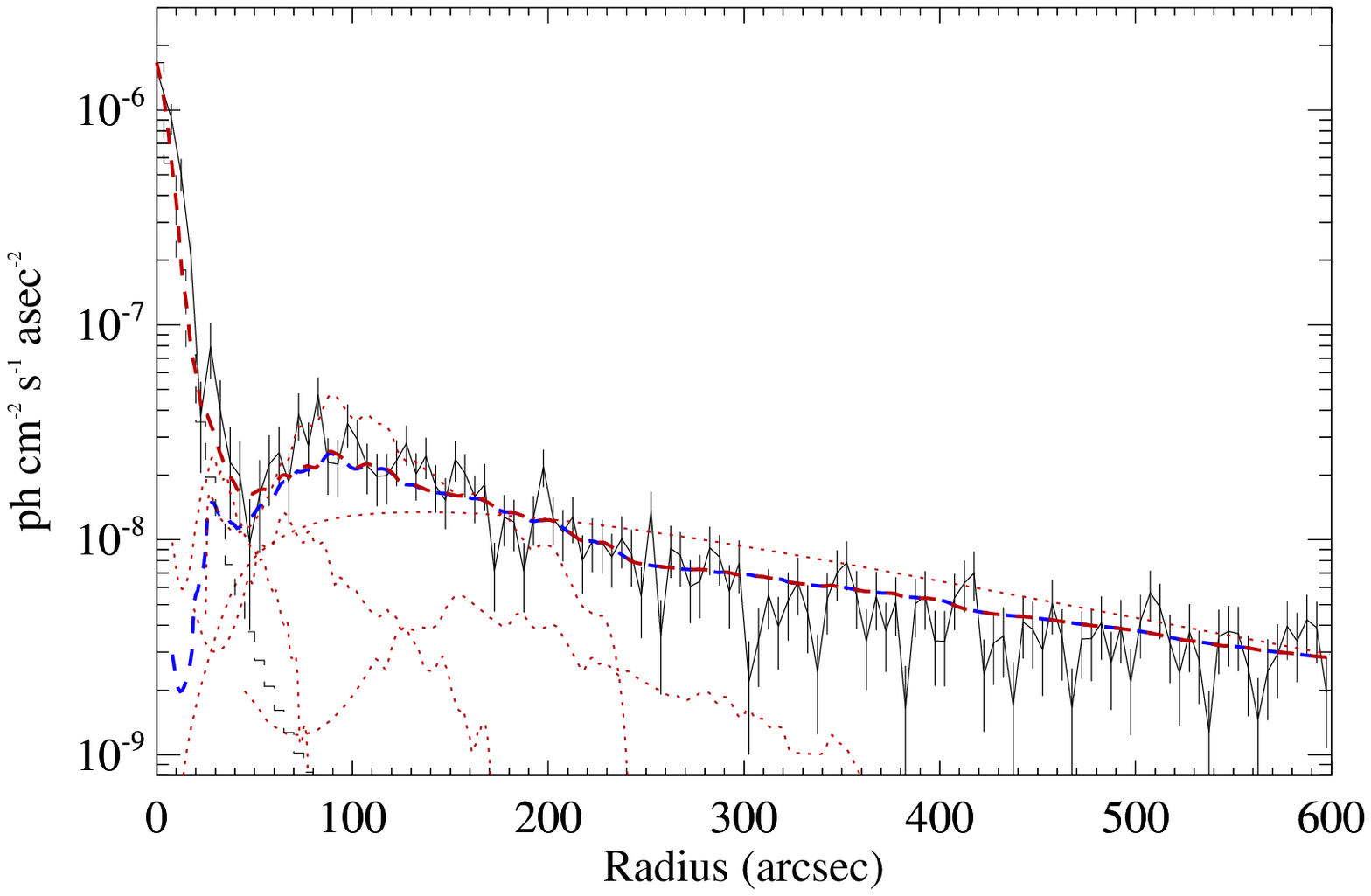}
\plottwo{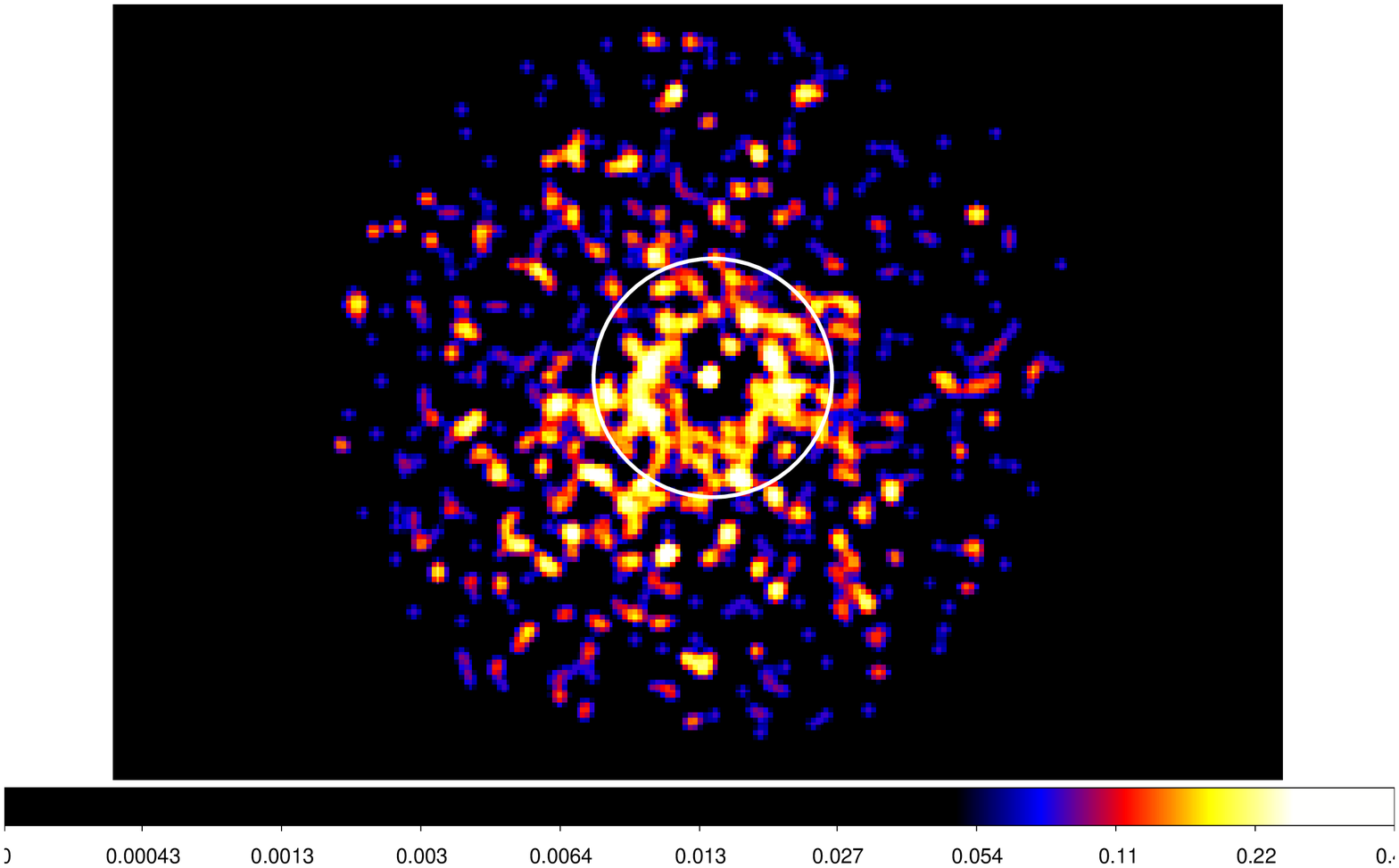}{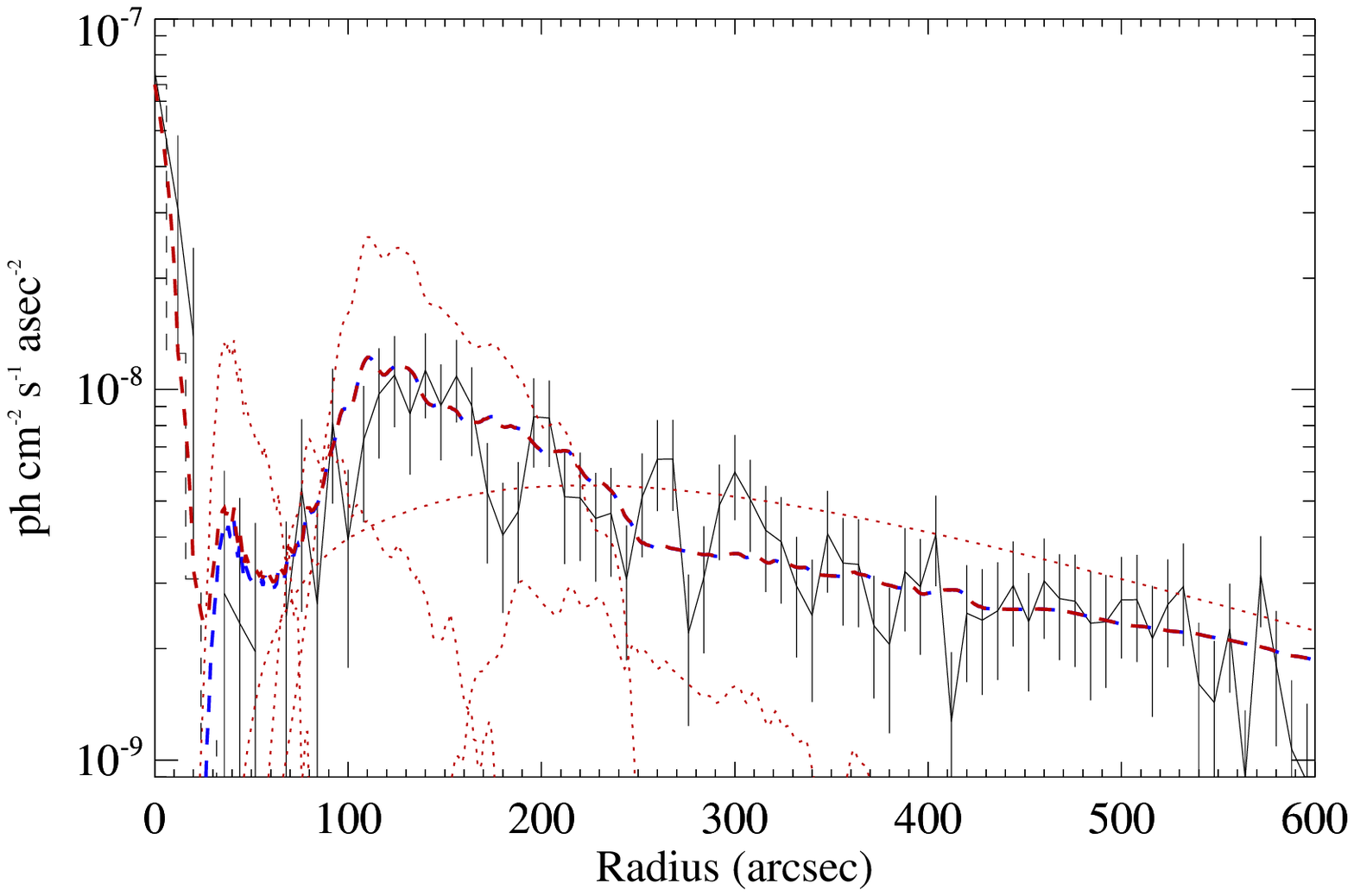}
\caption{\label{fig:swiftrings}
Left: Swift XRT images of 00031224047 (top) and 00031224052 (bottom) in 1.5-4.5 keV energy range smoothed to bring halo features out. The circular region has a radius of 240\arcsec. Right: SBP of the point source in 1.5-5 keV band and the DSH for the images on the left. Dashed lines indicate the XRT PSF \citep{Moretti09}. The dotted lines are contributions from each cloud before absorption. 
}
\end{figure*}

\swift\ SBP profiles for all pointing observations have been carried out in 1.5-4.5 keV band as described in \S\ref{subsec:dshmethod}. in Figure~\ref{fig:swiftrings}, we show the SBP profiles and modeling results of obsid 00031224047 which is the first PC mode observation during the 2016 decay, and obsid 00031224052 that shows the fully formed ring.  The ring formation is evident even for obsid 00031224047. Obsid 00031224006, the XRT observation which shows the anomalous soft state in T14, is discussed in detail in \S\ref{subsec:anomT14} The SBP models are discussed in \S\ref{subsec:sbpmodel}.

\section{Discussion}\label{sec:discussion}

The \chandra\ observation clearly shows a DSH ring with a profile such that there is a bright region between 80\arcsec\ and 250\arcsec\, with excess emission extending up to 600\arcsec\ (see Figures~\ref{fig:chanring} and \ref{fig:SBPdifen}). This ring profile is consistent with the overall evolution of the source in X-rays which shows a sharp rise in flux between MJD~57620 and MJD~57630, and slower decay in hard state after MJD~57740. \emph{Therefore, the most important result of this study is that there should be a massive dust cloud relatively close to the source as we see a single bright ring structure consistent with the outburst profile only days after the outburst ended.} As discussed in \S\ref{subsec:DSHintro}, the SBP together with the source flux history can be used to determine the distance to the source if the distance to the cloud is known. 

\subsection{Distance of \fu}\label{subsec:distance}

\begin{figure*}
\plottwo{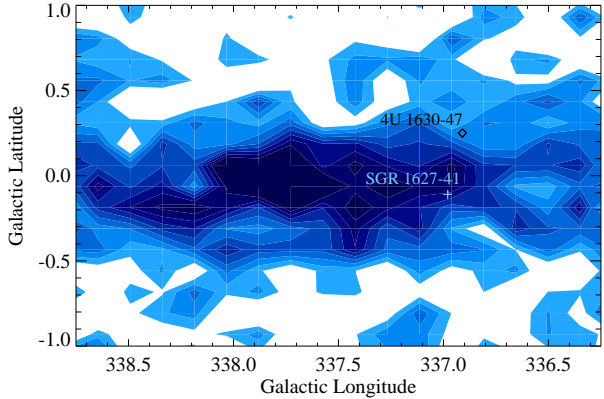}{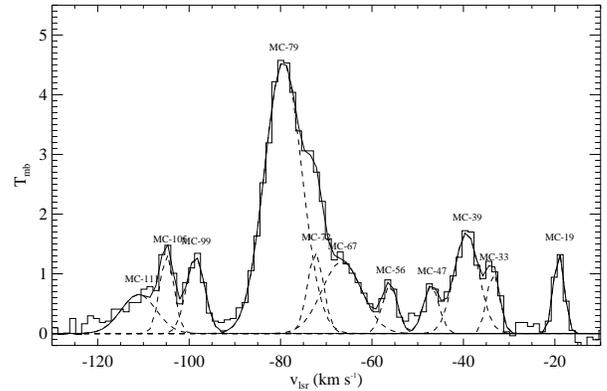}
\caption{\label{fig:integCO}
Left: Map of the integrated $^{12}$CO emission ($J=1-0$) using high resolution data of \cite{PlanckCO14}. The levels are spaced at 25 K km s$^{-1}$ starting at 50 K km s$^{-1}$. Right: $^{12}$CO $J=1-0$ spectrum towards the line of sight of \fu. The data is from \cite{Bronfman89}.
}
\end{figure*}

\fu\ is in a region of the Galaxy covered by a rich cluster of  \ion{H}{2} emitting clouds \citep{Mezger70}. In Figure~\ref{fig:integCO}, we show the integrated $^{12}$CO ($J=1-0$) velocity map from the \cite{PlanckCO14} data release showing the distribution of molecular clouds in the line of sight \citep[similar to Figure~2 in][showing both the SGR~1627$-$41 and \fu]{Corbel99}. The same Figure shows the spectrum towards \fu\ obtained using data from \cite{Bronfman89} CO survey. We fitted each feature in the spectrum with a Gaussian. We obtained near and far kinematic distances of all CO emitting clouds using the estimation based on the method described in \cite{Reid09} with the Galactic parameters from the model A5 in \cite{Reid14}. 

\epsscale{1.15}
\epsscale{1.0}

By comparing the optical extinction of \fu\ (derived from X-ray spectral fitting $N_{H}$ values) to the total optical extinction towards MC $-71$ as given in \cite{Corbel99},  \cite{Augusteijn01} argued that the source must be in front of the molecular cloud MC $-71$ which is located at 11 kpc. However, a close inspection of the position of \fu\ in the CO map (Figure~\ref{fig:integCO}, left) shows that at the position of the \fu, the integrated CO emission is much less than that of SGR~1627$-$41. Moreover, along the line of sight to \fu, the dominant emitter is not MC $-71$, but another cloud with velocity peaking at $-79$ km/s. The MC $-117$ - MC $-122$ complex is also not present in the line of sight to \fu. 

Since the DSH ring indicates one dominant scattering cloud we placed the source behind MC $-79$. Then, we obtained the distance to \fu\ using the SBP modeling, and double checked this result by recalculating the total extinction. To be able to do both of these calculations we need to resolve the ambiguity in kinematic distances of all clouds in the line of sight to \fu, especially the distance to MC $-79$.

\begin{figure}
\plotone{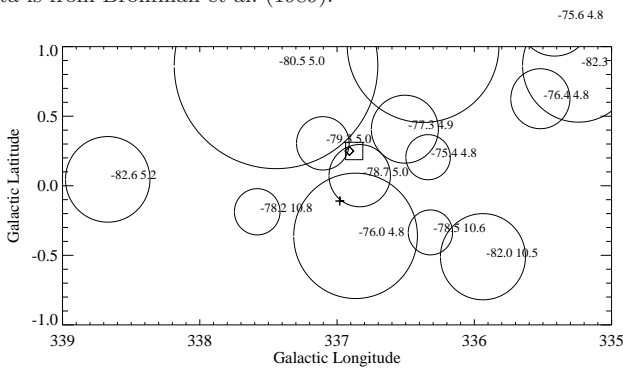}
\caption{\label{fig:cloud79v0}
Clouds with central velocity $\pm$ 4 km/s of MC $-79$ from \cite{MivilleD17} catalog. At the center of each cloud the velocity of the cloud is given. The second number is the distance determined by \cite{MivilleD17}. The diamond is the position of \fu, the cross is the position of SGR~1627$-$41 and the square is the resolution element of the \cite{Bronfman89} survey.
}
\end{figure}

Recent work by \cite{MivilleD17} identified 8107 molecular clouds in the entire \cite{Dame01} Survey of CO clouds between $b=-5^{\circ}$ and 5$^{\circ}$ utilizing Gaussian decomposition of the spectra. From this survey we identified clouds encompassing the position of \fu\ with central velocities within 4 km/s of the cloud velocities we show in Table~\ref{table:cloudpar}. An example map of clouds at central velocity close to $-79$ km/s is given in Figure~\ref{fig:cloud79v0}. To remove the ambiguity between the far and near distances, \cite{MivilleD17} compares the apparent radius of the cloud $R_{app}$ with the radius $R$ obtained using the correlation between line width $\sigma$ and $\Sigma R$, where $\Sigma$ is the surface density. A more common method is resolving distance ambiguity using the canonical $\sigma$ - $R$ correlation \citep{Solomon87}. In this work we adapted the $\sigma$ - $R$ correlation method. The results are given in Table~\ref{table:cloudpar}. 

For those cases where no cloud encompasses the source position (shown with stars in Table~\ref{table:cloudpar}), we assumed that these clouds are small and have not passed the pixel threshold in \cite{MivilleD17}. An inspection of the CO spectra within 0.5$^{\circ}$ of \fu\ from the \cite{Bronfman89} survey data supports this conclusion as the Gaussian components corresponding to MC $-105$, MC $-99$, and MC $-47$ are only present in a few pixels (7.5\arcmin $\times$ 7.5\arcmin resolution) around the source. Therefore we concluded that the apparent radii $R$ of these clouds (if they are real, and not a result of velocity crowding) are small. Using a comparison of $R$ with size from the $\sigma$ - $R$ correlation from the spectra shown in Figure~\ref{fig:integCO} we estimated that they are more likely to be at the near distance. The Gaussian component corresponding to MC $-33$ might be at the edge of  a cloud detected at \cite{MivilleD17} at -33.5 km/s which is estimated to be at the far distance. The $\sigma$ of MC $-111$ is high and therefore we could not resolve the ambiguity for the distance to this cloud. It is tentatively placed at the far distance.

\epsscale{1.2}
\epsscale{1.0}

Here, the most critical cloud is MC $-79$, and the ambiguity has not been resolved completely. The $\sigma$ - $\Sigma R$ relation indicates near distance, whereas $\sigma$ - $R$ relation indicates far distance. We favor the far distance from the SBP modeling, but for completeness we have analyzed the near-distance case as well.  

\begin{table*}
\centering
\caption{Cloud Parameters from the $^{12}$CO($J = 1–0$) Spectra along the Line of Sight to \fu}
\begin{tabular}{lcccccccc}
\hline \hline
Name & $V_{lsr}$\footnote{Local standard of rest velocity}  &  $\Delta$V(FWHM) & Near Dist. & Far Dist. & D ($\sigma$-$R$)\footnote{Distance ambiguity resolved based on $\sigma$-$R$ relation}  & D ($\sigma$-$\Sigma R$)\footnote{Distance ambiguity resolved based on $\sigma$-$\Sigma R$ relation} & $W(CO)$\footnote{Integrated $^{12}$CO emission ($J=1-0$)}   & $N(H_{2})$ \\
 & (km/s) & (km/s) & (kpc) & (kpc) & (kpc) & (kpc) & (K km/s) & (10$^{21}$ cm$^{-2}$) \\
\hline
MC $-111$ & -111.0 &  9.0 & 5.75 &  9.70 &  9.70* & - &6.3$\pm$0.5 &  1.2$\pm$0.4\\
MC $-105$ & -104.9 &  3.6 & 5.53 &  9.92 &  5.53* & - &5.1$\pm$0.8 &  1.0$\pm$0.3\\
MC $-99$ &  -98.6 &  4.7 & 5.32 & 10.15 & 5.32* & - &6.5$\pm$0.8 &  1.2$\pm$0.4\\
MC $-79$ &  -79.4 &  9.6 & 4.63 & 10.86 & 10.86 & 4.63 &46.2$\pm$1.7 & 8.8$\pm$2.6\\
MC $-72$ &  -72.3 &  4.1 & 4.37 & 11.14 & 11.14 & 11.14 &6.0$\pm$2.0 & 1.1$\pm$0.3\\
MC $-67$ &  -66.6 & 10.1 & 4.14 & 11.38 & 11.38 & 11.38 &13.0$\pm$3.0 &2.5$\pm$0.8\\
MC $-56$ &  -56.1 &  3.9 & 3.71 & 11.84 & 11.84 & 11.84 &3.3$\pm$0.7 & 0.6$\pm$0.2\\
MC $-47$ &  -47.0 &  4.0 & 3.30 & 12.27 & 3.30* & - &3.4$\pm$0.3 &  0.6$\pm$0.2\\
MC $-39$ &  -39.3 &  6.4 & 2.92 & 12.67 & 2.92 & 12.67 & 11.5$\pm$1.3 &  2.2$\pm$0.7\\
MC $-33$ &  -33.5 &  3.2 & 2.62 & 13.01 & 13.01* & - & 3.5$\pm$0.8 &  0.7$\pm$0.2\\
MC $-19$ &  -19.1 &  2.8 & 1.76 & 13.94 & 1.76 & 13.94 & 4.0$\pm$0.5 &  0.8$\pm$0.2\\
\hline
\multicolumn{9}{l}{* See text for the choice of distance.}
\label{table:cloudpar}
\end{tabular}
\end{table*}

\subsection{Distance to the source by modeling the dust distribution}
\label{subsec:sbpmodel}

\begin{figure}
\plotone{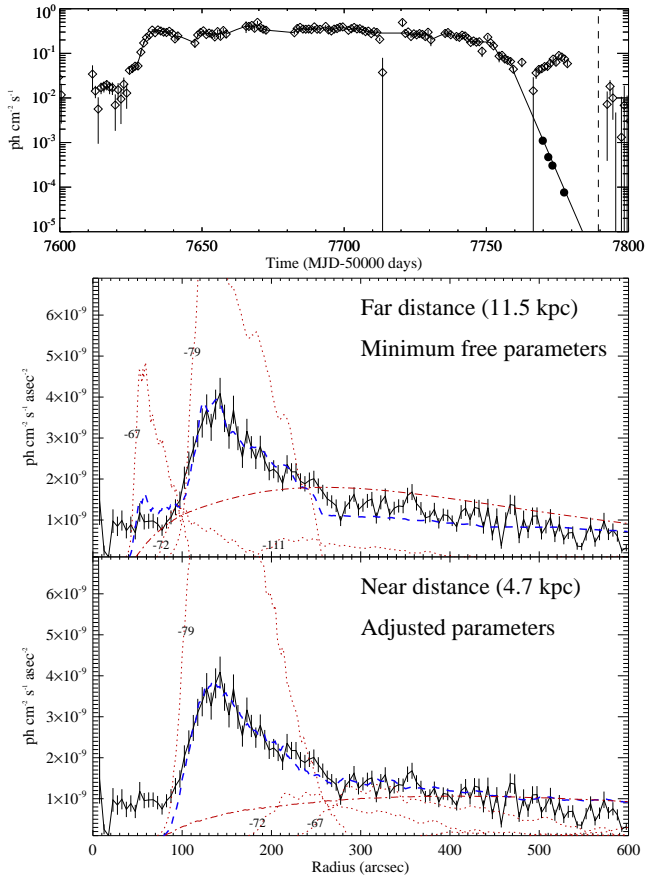}
\caption{\label{fig:profmodel}
Top panel: Flux evolution of the outburst in the 2-4 keV band. \maxi\ data are shown by diamonds, and \swift\ XRT data are shown with filled circles. The solid line is an extrapolated light curve using the \maxi and \swift\ data. Bottom two panels: SBP in 2.25-3.15 keV band and the model components. The dashed line is the overall model with absorption taken into account, the dotted lines are the contribution from each cloud before absorption and the dot dashed line is the contribution of a uniformly distributed dust in the continuum. See text for the distinction in far and near distance cases. Clouds are labeled according to their velocities.
}
\end{figure}

Standard geometrical analysis of scattering from a single dust cloud provides a relation between the time delay $\Delta t$, distance between the observer and the source $D$, distance to the dust layer $xD$ and the scattering angle $\theta$ (see Figure~\ref{fig:dshgeom}, and Equation~\ref{eqn:deltat}). Together with Equation~\ref{eqn:flux}, we can model the SBP which is a result of scattering of the source flux with dust present in different clouds. To be able to utilize Equation~\ref{eqn:flux} we need the entire flux evolution which we have thanks to \maxi\ and \swift\ monitoring. In the top panel of Figure~\ref{fig:profmodel} the 2-4 keV \maxi\  GSC light curve is shown. Beyond MJD~57750 the background measurement becomes unreliable. This can be clearly seen by the evolution of \swift\ XRT fluxes in the 2-4 keV band. Since the decay beyond MJD~57760 is well represented by an exponential decay, the entire flux history can be reconstructed. We also smoothed out sudden dips and peaks most probably caused by poor background subtraction in \maxi. To complete the analysis, we need the distances, $N_{H,r}$ content and thickness of all clouds between the source and the observer. For those clouds encompassing \fu\ in the \cite{MivilleD17} survey, we estimated the thickness based on the entire cloud size obtained using $R_{app}$ and the estimated distance as well as position of \fu\ within the cloud. For the rest, we again used the assumption that the $R_{app}$ must be small not to be detected in \cite{MivilleD17} and used generic 20 pc for near and 30 pc for far sources. 

\epsscale{1.2}
\epsscale{1.0}

To find the distance to the source we primarily match the rising part of the SBP in 2.25 - 3.15 keV band between 100" and 120" by adjusting the distance $D$ while keeping the main scatterer distance $xD$ at 10.86 kpc. The free parameters of our model are: a general normalization parameter (which adjusts for flux level differences in different energy bands as well as the constant $C$ in Equation~\ref{eqn:crosssec}), the thickness of the cloud MC $-79$ in the line of sight, the distance $D$, and the ratio of total $N_{H}$ to the total $N_{H}$ in clouds. The $N_{H}$ not in clouds is assumed to be distributed uniformly along the line of sight. We refer to this uniform component of the ISM as the 'continuum' hereafter. The total $N_{H}$ in the \chandra\ observation was determined to be $1.32 \times 10^{23}$ cm$^{-2}$. We also fitted the \maxi\ spectrum in the peak of the outburst with a model of $tbabs \times\ diskbb$ and obtained a similar value of $1.30 \times 10^{23}$ cm$^{-2}$. The $N_{H,r}$ in each dust cloud is found by taking the ratio of integrated velocities in CO spectra to that of MC $-79$ and normalizing such that the total is equal to the total $N_{H}$ in the clouds. We fixed $\alpha$ and $\beta$ to 3. Despite these simplifications, we were able to match the given SBP with this model by placing the source at 11.5 kpc. The individual contributions of dust clouds before absorption, and the overall fit is shown in Figure~\ref{fig:profmodel} in the panel labeled "Far distance, minimum free parameters".

As a way of checking our calculations, we have applied the same modeling without changing a single cloud parameter to earlier \swift\ observations. Only $\Delta t$ changes in Equation~\ref{eqn:flux}, and the agreement of the model with the data confirms our methodology (Figure~\ref{fig:swiftrings}).

\subsection{Validity of the approach and the distance error}

While the overall modelling represent the SBP in the most interesting region, there have been many simplifications applied, therefore it is important to understand the impact of these simplifications on the overall distance measurement.

\subsubsection{Ambiguity in cloud distance estimation and error in cloud distance estimation}

Resolving the ambiguity of cloud distances in the line of sight is not straightforward, and different approaches result in different distances. In our case, the distance of MC $-79$ is the critical measurement, and the near distance case will be discussed in \S\ref{subsec:neard}. As seen in Table~\ref{table:cloudpar}, there are other cases that the distances are ambiguous, but for those, the impact will be much lower as their $W(CO)$ indicates lower dust content, hence, their amplitudes are often low. 

MC $-67$ is the other important component with a strong impact on our conclusions.  First of all, if MC $-67$ is at the near distance our modeling will not be able to fit below 100\arcsec\ (but see \S\ref{subsubsec:chandra_back}) when MC $-79$ is at the far distance. 

At the far distance estimate of MC $-67$ favored by both $\sigma$ - $R$ and $\Sigma R$ - $R$ relation, the cloud is extremely close to the source, in fact, given the error in cloud distances, it is plausible that \fu\ is inside this cloud! With a small adjustment to its distance (well within the errors) and $N_{H,r}$ content, one can obtain a much better representation of the data. Similarly, while we had difficulty resolving the ambiguity of distance for MC $-111$, but placing it at the far distance, and a small adjustment of the $N_{H,r}$ of this cloud leads to a better representation of the data in our modeling. 

\subsubsection{Resolution of the sub-mm spectrum}

We have used the low resolution data of \cite{Bronfman89} for CO spectrum, yet the ISM is fractal (Corbel, S., private communication) and the integrated velocities in the line of sight to \fu\ can be significantly different. An example is the case of integrated velocity towards SGR~1627$-$41 obtained with 45\arcsec\ resolution \citep{Corbel99} which is significantly larger than that obtained from the\cite{Bronfman89} survey with a resolution of 7.5\arcmin. Using low resolution data does not change the fact that the source should be behind MC $-79$, however, the $N_{H,r}$ values in each cloud will change. Also, along the line of sight to \fu, the central velocity may be different introducing problems in determining the correct dust cloud. 

In Figure~\ref{fig:profmodel}, in the near distance case we adjusted the $N_{H,r}$ values of 3 clouds  and obtained a better representation of the entire SBP. With high resolution sub-mm observations we would have stronger constraints on the integrated velocity and therefore could fix $N_{H,r}$ values, and fitting the data then would require adjusting their thicknesses, and dust properties.

\subsubsection{Modeling of the \chandra\ background SBP}
\label{subsubsec:chandra_back}

One of the arguments for MC $-79$ being at the far distance is that it places the source very close to MC $-66$ and therefore we could model the SBP below 100\arcsec. If we underestimated the background SBP in the \chandra\ data, and the excess between 20\arcsec -- 100\arcsec\ is absent, then the distance ambiguity will be harder to resolve (see \S\ref{subsec:neard}). Apart from this, a possible incorrect estimation of the background SBP will also impact the ratio of $N_{H}$ in dust clouds and the continuum significantly because the profile beyond 300\arcsec\ is almost entirely due to dust in the continuum in our model. Its effect on the distance estimate will be small if the background SBP is constant at the region of interest.

On the other hand, since the same model fits the \swift\ XRT SBP as well, if we are underestimating \chandra\ background in the SBP, we should also be underestimating \swift\ XRT background in the SBP.

\subsubsection{The X-ray flux evolution}
\label{subsubsec:fluxevol}

The flux evolution has been determined using \maxi\ and \swift. Yet, the region between MJD 57750 and MJD 57770 bear some uncertainty as this is the region \maxi\ background is unreliable and no \swift\ observations were possible because of the Sun angle constraint. This region corresponds to 80\arcsec -- 120\arcsec\ region in the SBP. The exponential decay approximation works well in terms of representing multiple \swift\ observations. 

We also have not done a correction to \maxi\ fluxes due to the dust scattering halo. We can justify this as follows. On MJD~57772 the flux in the PSF of \swift\ XRT is comparable to the flux in the DSH. We stop using \maxi\ data on MJD~57761, and the overall \maxi\ flux is a factor of 70 larger than the flux on 
MJD~57772, whereas the SBP modeling indicates only a factor of 3 more flux in the DSH. Therefore for the data range we use the \maxi\ data, the DSH contribution to \maxi\ is negligible.

\subsubsection{Other parameters}

The other notable parameters that can impact the overall model are variations of scattering cross section per hydrogen atom, $\alpha$, and $\beta$ parameters in the clouds. These will result in minute differences in the overall shape of the model. The $\alpha$ parameter will also effect the ratio between the $N_{H}$ in the clouds and continuum because clouds at lower scattering angles will be affected more strongly with $\alpha$. 

The quality of the SBP data is not high enough to differentiate these minute differences over the more important effects. One can envision several ways to improve the distance measurement with this method, as well as constraining cloud distances and dust properties by even fitting fine features in the SBP: 

\begin{itemize}
	\item Better quality \chandra\ data taken at a higher luminosity. The observation must be taken at a level in outburst decay such that the ring has formed and the SBP of the cloud and the PSF are not intersecting.
	\item High resolution sub-mm data to obtain the cloud velocities and distances more accurately. This will not only help in resolving ambiguity but also provide more accurate $N_{H,r}$ values that can be fixed allowing other parameters to be varied.
	\item \swift\ XRT (or any other low energy imaging instrument) monitoring at times \maxi\ has problems with background. 
	\item Detailed spectral modeling of the dust scattering halo and comparing dust properties obtained from spectral modeling to expected parameters.
	
\end{itemize}

\subsection{The alternative distance estimate}
\label{subsec:neard}

 As discussed earlier, MC $-79$ could also be at a distance of 4.63$\pm$0.25 kpc. We have adjusted the model parameters to model the SBP for the case that MC $-79$ is at the near distance (see Figure~\ref{fig:profmodel}, labeled "Near distance, Adjusted parameters"). To be able to get a meaningful fit we needed to place all clouds in the near distance estimate.  In this case we get the best fit at a distance of 4.7 kpc. In the near distance case, we cannot explain the continuum SBP below 100\arcsec\ unless our background estimate is wrong, or, MC $-79$ is not a single cloud but consist of a couple of clouds (a larger cloud with a central velocity of \wsim $-79$ km s$^{-1}$ and a smaller cloud at \wsim $-81$ km s$^{-1}$). This scenario is not ruled out completely, as the modelling in \cite{MivilleD17} indicates that two clouds with velocities $-78.7$ km s$^{-1}$ and $-79.3$ km s$^{-1}$ encompass the position of \fu, and another cloud with a central velocity of $-80.5$ km s$^{-1}$ extends to the edge of the the resolution element that includes \fu, and all these clouds are at the near distance according to the $\sigma - \Sigma R$ correlation distance estimate (see Figure~\ref{fig:cloud79v0}). 
 
 Moreover, with the near distance estimate, it is difficult to model the SBP beyond 200\arcsec as well. To be able to get a fit we placed MC $-67$ and MC $-72$ at the near distance, but for both clouds, both methods of resolving distance ambiguity points to the far distance estimate. Removing contributions from these clouds (see Figure~\ref{fig:profmodel}) results in a big gap between 200\arcsec\ and 400\arcsec\ that cannot be filled with a simple adjustment of the continuum parameters.

\subsection{Extinction towards \fu}

Now we can check whether the $N_{H}$ values we use in our modeling and those found in general in the literature are consistent with the extinction estimated from CO maps towards the line of sight to \fu.
The integrated spectrum for all clouds in front of \fu\ for MC $-79$ at the far distance, $W(CO)$=102 K km s$^{-1}$.  Using the conversion factor $X_{CO} = N(H_{2})/W(CO)$ of ($2 \pm  0.6$) $\times$ 10$^{20}$ molecules cm$^{-2}$ (K km s$^{-1}$)$^{-1}$ \citep{Bolatto13} we found a molecular hydrogen column density N(H$_{2}$) of (20.4$\pm$6.1) $\times$ 10$^{21}$ cm$^{-2}$. The total hydrogen column density is $N(H)$ = $N(H_{I})$ + 2$N(H_{2}$). Using $H_{I}$ surveys, \cite{Augusteijn01} cites a contribution of 2 $\times$ 10$^{22}$ cm$^{-2}$ to the total $N_H$. Since this could be an underestimation, the lower limit to total $N_{H}$ is then estimated to be 6.1 $\times$ 10$^{22}$ cm$^{-2}$ towards the line of sight of \fu. The X-ray derived $N_{H}$ values are often much larger than 6 $\times$ 10$^{22}$ cm$^{-2}$ \citep[this work, T14,][]{Wang16}. Therefore, the extinction determined from the radio maps when the source is behind MC $-79$ is consistent with extinction derived using X-ray spectral fitting. In our modeling, we took the total $N_{H}$ to be 13 $\times$ 10$^{22}$ cm$^{-2}$, and best representation of the data is obtained when the distribution of dust in the molecular clouds and in the continuum is 52\% to 48\% which is consistent with the $N_{H}$ in clouds derived from the integrated velocity from \cite{Bronfman89} survey.

Finally, using the conversion factor of $N_{H} = (2.87 \pm 0.12) \times 10^{21} A_{V}$ cm$^{-2}$ \citep{Foight16} which uses the ISM abundances of \cite{Wilms00} and the XSPEC $tbabs$ absorption model as used in this work, we obtain an optical extinction of  45.

\epsscale{1.2}
\begin{figure}
\plotone{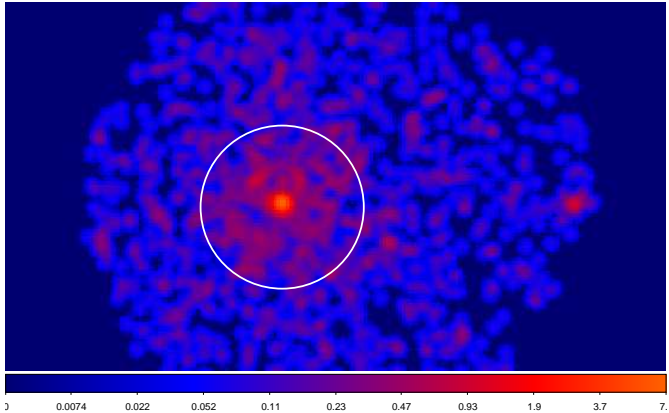}
\caption{\label{fig:halo06}
Swift XRT image of obsid 00031224006 in 1.5-5 keV energy range smoothed to bring halo features out. The circular region has a radius of 240\arcsec.
}
\end{figure}
\epsscale{1.0}

\subsection{Anomalous soft state revisited}
\label{subsec:anomT14}

One of the motivations for this work was to assess whether the anomalously low luminosity soft state observation during the 2010 decay of \fu\ has been an artefact of a dust scattering halo softening the point source spectrum. In fact, a DSH is present in the XRT image of obsid  0003124006 (see Figure~\ref{fig:halo06}), and the 47\arcsec\ source region used in T14 could include softer emission from the halo. In this section we try to assess how much of the emission is from the DSH within the 47\arcsec\, and when this contribution is taken into account, whether the actual spectrum of the point source is hard. 

\begin{figure}
\plotone{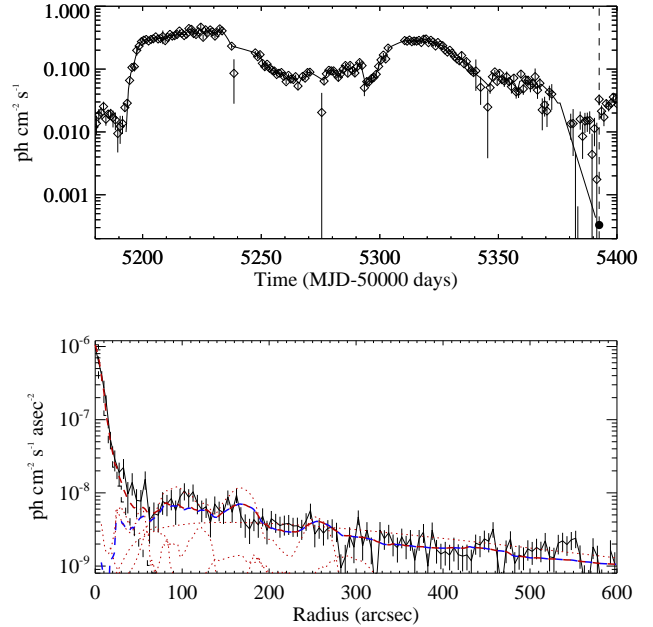}
\caption{\label{fig:swift06prof}
Top: Flux evolution of the 2009 outburst of \fu\ in the 2-4 keV band, \maxi\ data are shown by diamonds, and \swift\ XRT flux is shown with filled circle. The solid line is extrapolated light curve using the \maxi\ and \swift\ data. Bottom: SBP and the model components. The blue dashed line is the overall DSH contribution, and the dotted lines are contribution from each cloud. The red dashed line is the DSH + PSF. 
}
\end{figure}

Based on the \chandra\ SBP modeling we already obtained possible distribution of dust clouds and thicknesses along the line of sight. To verify methodology and for consistency check we applied exactly the same model without changing any parameter to the SBPs of \swift\ observations on MJD~57771.8 and MJD~57781.9 and obtained the results shown in Figure~\ref{fig:swiftrings}. The model represents the data well. Then we took the same dust distribution and same form of the cross section, and applied it for obsid 0003124006, again without tweaking any parameters (even though now it is a different outburst where the spectral evolution is slightly different than the 2016 outburst). Despite the simplicity of our approach the modeled SBP represents the data well as shown in Figure~\ref{fig:swift06prof}. The SBP model shown indicates that the dust scattering halo is not affecting the 47\arcsec\ spectral region used in T14, and indeed that this observation is anomalously soft. We note that, if the source is at the near distance estimate, the anomolous soft state observation would have been at a factor of four lower Eddington luminosity fraction given in T14. 

\epsscale{1.2}
\epsscale{1.0}

\subsection{Nature of \fu}\label{subsec:nature}

While many of the X-ray spectral and timing properties of this source is similar to other low mass X-ray binaries (LMXB), there also are arguments in favor of this system being a high mass X-ray binary (HMXB). \cite{Augusteijn01} point out that donor star should be intrinsically luminous, because the amplitude of the variation in the IR from quiescence to outburst was small. It might also help explain why the rise time of the outburst is short, and it might explain the high duty cycle of the outbursts, and why they are fairly erratic (as seen in Figure~\ref{fig:maxiasm}). 

For the far distance scenario, the SBP between 20\arcsec and 80\arcsec can be modeled by a dust cloud (MC $-66$) very close ($<$100 pcs) to \fu. For the near distance scenario, the modeling indicates that \fu\ should only be 70 pc from MC $-79$. If we assume that the X-ray binary was produced in the nearby cloud,  and that it was born with a low natal kick \citep[e.g. 30 km/s, at the low end of kick velocities discussed in][]{MillerJ14}, it would take approximately 3 Myr to get to 100 pc distance. A potential problem with this timescale  is whether the source would have time to evolve into Roche lobe overflow, and whether the system would be circular enough that we wouldn't see an obvious orbital period signature.  This would not be a problem if the source is a HMXB. 

The Be X-ray binary scenario gives a reasonable way to make the system so close to a molecular cloud without it being a coincidence.  This allows the system to show strong mass transfer without having to re-circularize and get it back into Roche lobe overflow in \wsim\ 2 Myr.  It also gives a way to have rapid rises to the outbursts.  It might also explain the "double" outbursts that are sometimes observed, if there are two passages through the decretion disk near the periastron for those outbursts, and both transfer a significant amount of mass. This system shows a mixture of normal outbursts which are nearly periodic, and then much stronger outbursts as seen in Figure~\ref{fig:maxiasm}.  This could be explained as Type I and Type II outbursts for a Be X-ray binary \citep{Reig11, Martin14}.  

However, the outbursts are not exactly periodic and are complex \citep{Kuulkers97} and may not be related to the orbital period as expected from Be binary systems. Alternatively, if the system is a low mass X-ray binary, it may have an evolved donor star in order to explain the low amplitude of variability in the infrared.  In such a case, the variations on the brightness and durations of the outbursts may be the result of the tidal-thermal instability producing "super-outbursts" \citep{Osaki96}.  The two scenarios can be distinguished in future work by making high cadence infrared variability measurements.  In the Be X-ray binary scenario, the normal outbursts proceed on a periodicity which is the orbital period.  In the superoutburst scenario, the orbital period will be considerably shorter than the time separation between outbursts, and there should be modulations due to superhumps which will be at a period within a few percent of the orbital period \citep[e.g.][]{Haswell01}.

\section{Summary and future work}

We have analyzed the dust scattering halo of \fu\ observed with \chandra\ and \swift\ during the decay of its 2016 outburst. The ring structure of the SBP indicates that the source is behind a massive molecular cloud which is determined to be MC $-79$ from CO maps. While the kinematic distance estimate of MC $-79$ is ambiguous, we favor the far distance estimate because it allows us to model the SBP less than 100\arcsec better.  

The ambiguity in the source distance can be resolved with a more detailed spectral analysis of the halo which we plan to conduct in a future work. Also, higher luminosity \chandra\ observations and a higher resolution sub mm mapping towards the line of sight would also provide additional information to break the degeneracy.

Given the large number of clouds along the line of sight, the probability of a source lying quite close to a cloud is relatively high.  There are uncertainties due to the near-far degeneracy, but it is likely that more than 10\% of the line of sight is within 500 pc of a molecular cloud.  Still, an association with a molecular cloud could, potentially, help explain several of the peculiarities of this source.  First, a cloud might provide a "working surface" for the jet to interact with that would not be present for most soft X-ray transients.  This could, in turn, explain the baryonic content of the jets found from gratings observations.  Second, \cite{Augusteijn01} found that the ratio of the amplitude of the infrared variability to the X-ray variability was anomalously low for an low mass X-ray binary.  This could be explained by having a high mass donor star, in association with the star formation in the cloud.

Normally, high mass X-ray binaries are not transients unless the donor star is a Be star.  A Be donor might explain a few additional elements of the phenomenology of the outbursts of this system.  The outbursts are nearly (although not strictly) periodic, except for some very strong outbursts.  These may be indicative of Type I/Type II outbursts from Be transients.

A clear observational test exists for determining whether the HMXB scenario is a reasonable one.  If the system was formed in a particular molecular cloud and stayed there, it should have a space velocity very close to the space velocity of the cloud.  The radial velocity can be determined from infrared emission lines in outburst.  The proper motion can be determined, eventually, by very long baseline interferometry radio measurements taken during the outburst.


\acknowledgments E.K, acknowledges T\"UB\.ITAK 1001 Project 115F488.  JAT acknowledges partial support from an XMM Guest Observer grant for studying the 4U 1630-47 low soft state and Chandra grant GO5-16152X for studying hard X-ray transients in the Galactic Plane. Support for this work was provided by the National Aeronautics and Space Administration through Chandra Award Number GO7-18040X issued by the Chandra X-ray Center, which is operated by the Smithsonian Astrophysical Observatory for and on behalf of the National Aeronautics Space Administration under contract NAS8-03060. We thank \chandra\ flight team and especially Director Wilkes for their efforts to schedule the \chandra\ observation in a manner that maximized the scientific return. This research has made use of \maxi\ data provided by RIKEN, JAXA and the \maxi\ team. We thank the Neil Gehrels Swift Observatory MOC for scheduling TOO observations that enabled this work. We thank Chris Nixon, Stephane Corbel, Tom Russell, Sergey Tsygankov, Tolga G{\"{u}}ver, Yuki Kaneko G{\"{o}}{\u{g}}{\"{u}}{\c{s}}, Ersin G{\"{o}}{\u{g}}{\"{u}}{\c{s}} and Sel\c{c}uk Bilir for useful discussions.  


\bibliographystyle{jwapjbib}





\end{document}